\documentclass[preprint2]{aastex}

\slugcomment{Not to appear in Nonlearned J., 45.}

\shorttitle{Which galaxies host bars and disks?}
\shortauthors{M\'endez-Abreu et al.}

\begin{document}


\title{Which galaxies host bars and  disks? \\
       A study of the Coma cluster}


\author{J. M\'endez-Abreu\altaffilmark{1,2} and R. S\'anchez-Janssen\altaffilmark{3} and J. A. L. Aguerri\altaffilmark{1,2}}
\email{jairo@iac.es, rsanchez@eso.org, jalfonso@iac.es}


\altaffiltext{1}{Instituto de Astrof\'isica de Canarias, Calle V\'ia L\'actea s/n,
  E-38200 La Laguna, Tenerife, Spain}
\altaffiltext{2}{Departamento de Astrof\'isica, Universidad de La Laguna, 
              E-38205 La Laguna, Tenerife, Spain}
\altaffiltext{3}{European Southern Observatory, Alonso de Cordova 3107, Vitacura, Santiago, Chile}


\begin{abstract}
We present  a study of the  bar fraction in the  Coma cluster galaxies
based  on a  sample of  $\sim 190$  galaxies selected  from  the Sloan
Digital Sky  Survey Data  Release 6 (SDSS-DR6)  and observed  with the
Hubble Space  Telescope (HST) Advanced  Camera for Survey  (ACS).  The
unprecedented resolution of the HST-ACS images allow us to explore the
presence  of bars,  detected  by visual  classification, throughout  a
luminosity  range  of 9  magnitudes  ($-23  \lesssim$ M$_{r}  \lesssim
-14$), permitting us to study the poor known region of dwarf galaxies.
We find  that bars  are hosted by  galaxies in  a tight range  of both
luminosities ($-22 \lesssim$ M$_{r} \lesssim -17$) and masses ($10^{9}
\lesssim {\cal M_{*}}/{\cal M}_{\sun} \lesssim 10^{11}$).  This result
holds when  comparing with a sample of  bright/massive field galaxies.
In addition, we find that the bar fraction does not vary significantly
when going  from the  center to the  cluster outskirts,  implying that
cluster   environment    plays   a   second   order    role   in   bar
formation/evolution.
The  shape of  the  bar  fraction distribution  with  respect to  both
luminosity and mass is well  matched by the luminosity distribution of
disk galaxies in  Coma, indicating that bars are  good tracers of cold
stellar disks.
We  discuss the  implications of  our  results for  the formation  and
evolution scenarios of bars and disks.
\end{abstract}





\section{Introduction}
\label{sec:intro}

Bars are  believed to be very  important with regard  to the dynamical
and secular evolution of  disk galaxies. In particular, they represent
the main internal driver  of galaxy structure and morphology evolution
within the central $\sim$10 kpc.   Stellar bars are also recognized as
the  most  important internal  factor  that  redistribute the  angular
momentum   between   the   baryonic   and   dark   matter   components
\citep{debattistasellwood98,debattistasellwood00}.    The   amount  of
angular momentum  exchanged is related  to specific properties  of the
galaxies,  such as  the  bar  mass, halo  density,  and halo  velocity
dispersion \citep{athanassoula03,sellwooddebattista06}. Moreover, they
funnel material towards the  galaxy center where starbursts can ignite
\citep{sheth05}, contribute  to the formation  of bulge-like
structures  \citep[][]{kormendykennicutt04}, inner  star-forming rings
\citep[][]{buta03,munoztunon04},  inner bars  \citep[][]{erwin04,debattistashen07}, and
feed   the   central   black  hole   \citep[][]{shlosman00,corsini03}.
Peanut/boxy bulges in galaxies are  also thought to be associated with
bending     instabilities      and     bar     vertical     resonances
\citep{bureaufreeman99,martinezvalpuesta06,mendezabreu08}

Stellar bars are observed in optical images of roughly half of all the
nearby disk galaxies  \citep{barazza08,aguerri09}. This fraction rises
slightly  to  about 59-62\%  when  near-infrared  images are  analysed
\citep{laurikainen04,marinovajogee07,   menendezdelmestre07}.   It  is
established that  they appear naturally in most  simulations of galaxy
formation once  a dynamically cold and  rotationally-supported disk is
at place. However, even if bars  are ubiquitous in the universe, it is
not clear yet why one galaxy can exhibit a bar structure while another
apparently similar does not.

The mechanisms  leading to the formation  of bars can  be divided into
internals and  externals. The most widely  accepted internal mechanism
to  produce  bars  in galaxies  is  based  on  the $m=2$  mode  global
instability     in      cold,     rotationally     supported     disks
\citep{hohl71,ostrikerpeebles73}.  Environmental effects could also be
important,  though  the  existence  of  competing  mechanisms  usually
prevents   simple   interpretations.    Depending   on   the   orbital
configuration, interactions can weaken  severely bars and even destroy
them  or, on  the other  hand,  significantly speed  up bar  formation
\citep{noguchi87,aguerrigonzalezgarcia09}.      Moreover,    numerical
simulations  of galaxy  harassment  in clusters  have  shown that  bar
growth  is  a  common  and  stable process  during  the  evolution  of
late-type         galaxies        in         dense        environments
\citep[e.g.][]{mastropietro05}.

Only with  the recent advent of  large galaxy surveys,  either at high
\citep[][COSMOS]{sheth08}           or           low          redshift
\citep[][SDSS]{barazza08,aguerri09},   statistical   studies  of   bar
frequencies  have  been  possible.   However, bar  studies  have  been
usually  restricted to  luminous galaxies  due to  either the  lack of
spatial resolution or because images were not deep enough. The present
work attempts to put  observational constraints on the internal (mass)
and external (environment) parameters  that influence bar formation by
carrying out  a comprehensive  study of the  bar fraction in  the Coma
cluster  galaxies throughout a  wide range  of 9  magnitudes, covering
from giant  ellipticals (M$_{r} \sim  -23$) to dwarf  galaxies (M$_{r}
\sim -14$).   This research will  also provide us  the luminosity/mass
interval where cold  stellar disks are present in  galaxies.  For this
purpose we take advantage of  the unrivalled resolution of the HST-ACS
Coma cluster Treasury Survey, which  provides deep imaging of the core
and infall region of the Coma cluster.

The paper is  organised as follows. The galaxy sample,  as well as the
selection of  the Coma cluster  member galaxies is presented  in Sect.
\ref{sec:data};  the method adopted  to detect  bar structures  in the
sample  galaxies, and  the  results obtained  are  explained in  Sect.
\ref{sec:results}; the  discussion of the results  and our conclusions
are  given in  Sect. \ref{sec:conclusions}.  Throughout the  paper, we
assume a distance modulus of m-M=35.

\section{Data and cluster membership selection}
\label{sec:data}

The Coma cluster is one of the best-studied galaxy clusters because of
its  relative  proximity  and  because   it  may  be  a  prototype  of
dynamically relaxed cluster, even though significant substructures are
present \citep{collessdunn96}.

The  HST-ACS  Coma Cluster  Treasury  Survey \citep{carter08},  covers
$\sim$ 230 arcmin$^2$ with 21 ACS pointings ($\sim$3$\times$3 arcmin).
The magnitude  limits of  the survey at  10 $\sigma$ for  1 arcsec$^2$
extended  regions  are  g'=25.8  mag/arcsec$^2$ and  $I_{\rm  C}$=25.0
mag/arcsec$^2$.  At the distance of the Coma cluster ($\sim$ 100 Mpc),
the resolution  of HST-ACS ($0\farcs1$)  corresponds to $\sim  50$ pc.
This gives  essentially the  same physical resolution  as ground-based
observations have in  Virgo and it will allow us  to resolve bars down
to sizes of $r_{\rm bar}\sim 150$ pc.

The  Coma  cluster is  also  covered  by  the SDSS,  providing  galaxy
magnitudes in five bands ($u, g ,r, i, z$). For the sake of comparison
with recent works on bar fractions \citep{barazza08,aguerri09}, and in
order to have access to galaxy colors, which will help us to determine
cluster  memberships, we decided  to create  our catalogue  of sources
using the SDSS-DR6 \citep{adelman08}.

\begin{figure*}[!ht]
  \centering
  \includegraphics[width=\textwidth]{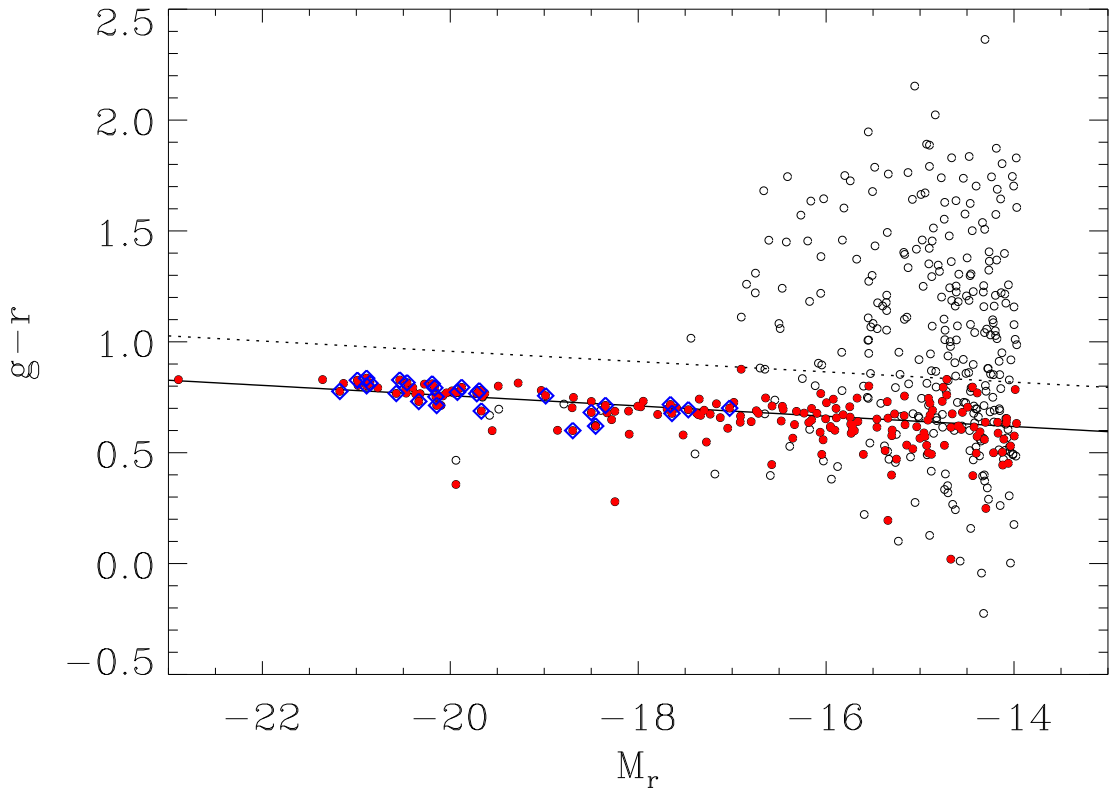}
  \caption{Color-magnitude  diagram  for our  complete  sample of  477
    galaxies (black circles). Red points represent the subsample of 188
    galaxies  considered as members  of the  Coma cluster,  while blue
    diamonds point out those Coma galaxies which host a bar. The solid
    and dotted  black line shows our  fit to the red  sequence and our
    limit   of  0.2   magnitudes  above   the  fitted   red  sequence,
    respectively.}
              \label{fig:cm}%
  \end{figure*}

The steps followed to obtain  our final sample of Coma cluster members
were the  following: from  the SDSS-DR6 we  downloaded a  catalogue of
extended sources within  a 5 arcmin radius from  the position of every
ACS pointing.   This catalogue contains all  galaxies with m$_{r}<21$,
which  represent  approximately the  completeness  limit  of the  SDSS
photometric survey  for extended sources, and with  $b/a>0.5$, $a$ and
$b$ being the semi-major and  semi-minor axis lengths of the galaxies,
in order to  deal with projection effects.  This  resulted in a sample
of 477  galaxies (black  circles in Fig.   \ref{fig:cm}), 104  of them
having  recession  velocities available  from  the Nasa  Extragalactic
Database (NED).   We select galaxies  with velocities $\pm  3000$ km/s
with  respect to  the Coma  redshift  as cluster  members. This  range
corresponds to  a 3$\sigma$  cut on the  velocity distribution  of the
Coma cluster galaxies ($\sigma \sim 1000$ km/s; Colless \& Dunn 1996).
We found  that all 104  galaxies satisfy this condition  and therefore
they  are cluster  members.  At  this  point, we  decided to  visually
inspect  every  galaxy in  order  to  determine  its possible  cluster
membership based on its morphology.  We follow the prescriptions given
by  \citet{michardandreon08}  where  they claim  that  morphologically
selected  cluster members  are  reliable when  compared with  redshift
membership,  and present  a  catalogue  of 473  Coma  members down  to
M$_{B}$=-14.25 based on their morphology. The procedure to distinguish
between background objects and cluster  members is based in some basic
hypothesis.  Typical  spirals in the  background lie in  the magnitude
range of dwarf  galaxies in Coma.  However, these  objects are rare in
the field and  even more in dense cluster  environments.  Moreover, we
exclude from our sample all spiral like galaxies that had either bulge
or disk sizes too small if  they were cluster members.  With regard to
elliptical  galaxies, far-away bright  ellipticals are  generally more
concentrated that  early-type dwarfs in  Coma.  In addition,  they are
more reddened and  our further color cut will easily  get rid of them.
We found that,  from the remaining 373 galaxies  without redshift, 127
galaxies followed  the morphological  criteria to be  cluster members,
while 246 did  not.  A further colour condition  was still imposed for
these candidates to be considered as cluster members. To calculate our
color cut, we  fit the red sequence of all galaxies  in our sample and
imposed  that  members  should  have  a  $g-r$  color  less  than  0.2
magnitudes above the  value of the fit (see  Fig. \ref{fig:cm}).  With
this  further constraint,  our  sample of  Coma  {\it secure}  members
consists  of 188  galaxies  (red points  in  Fig.  \ref{fig:cm})  with
magnitudes in the range $-23 <$ M$_r < -14$.

Several quality  checks were done during  the morphological membership
classification:  at  a first  step  the  classification was  performed
independently  by  two  of   the  authors  (JMA  and  RSJ),  obtaining
consistent  results.  Next  we compared  our cluster  memberships with
those  given by  \citet{michardandreon08}.  From  the 225  galaxies in
common,  both  classifications  agree  for  a 78\%  of  galaxies.   In
addition, including  colors at  this step instead  of only as  a final
constraint, this  percentage grows up  to a 85\%, confirming  the good
agreement  between both classifications.   As a  final check,  we also
compared our morphological membership with the available spectroscopic
redshifts.  We  found that  from the 104  galaxies with  redshift, 103
(99\%)   share  the  same   class,  giving   validity  again   to  our
classification.

\section{Method and results}
\label{sec:results}

The presence  of a  bar can  be revealed by  visual inspection  of the
images \citep{devaucouleurs91}, by analysing the shape and orientation
of                 the                 galaxy                isophotes
\citep[][]{laurikainen05,menendezdelmestre07,marinovajogee07,sheth08,
  barazza08,aguerri09,marinova09},  by studying  the Fourier  modes of
the                         light                         distribution
\citep[][]{ohta90,elmegreenelmegreen85,aguerri98,aguerri00,laurikainen05},
or  fitting  the  different   structural  components  to  the  surface
brightness distribution \citep[][]{weinzirl09}.

\begin{figure*}[!ht]
  \centering
  \includegraphics[width=0.49\textwidth]{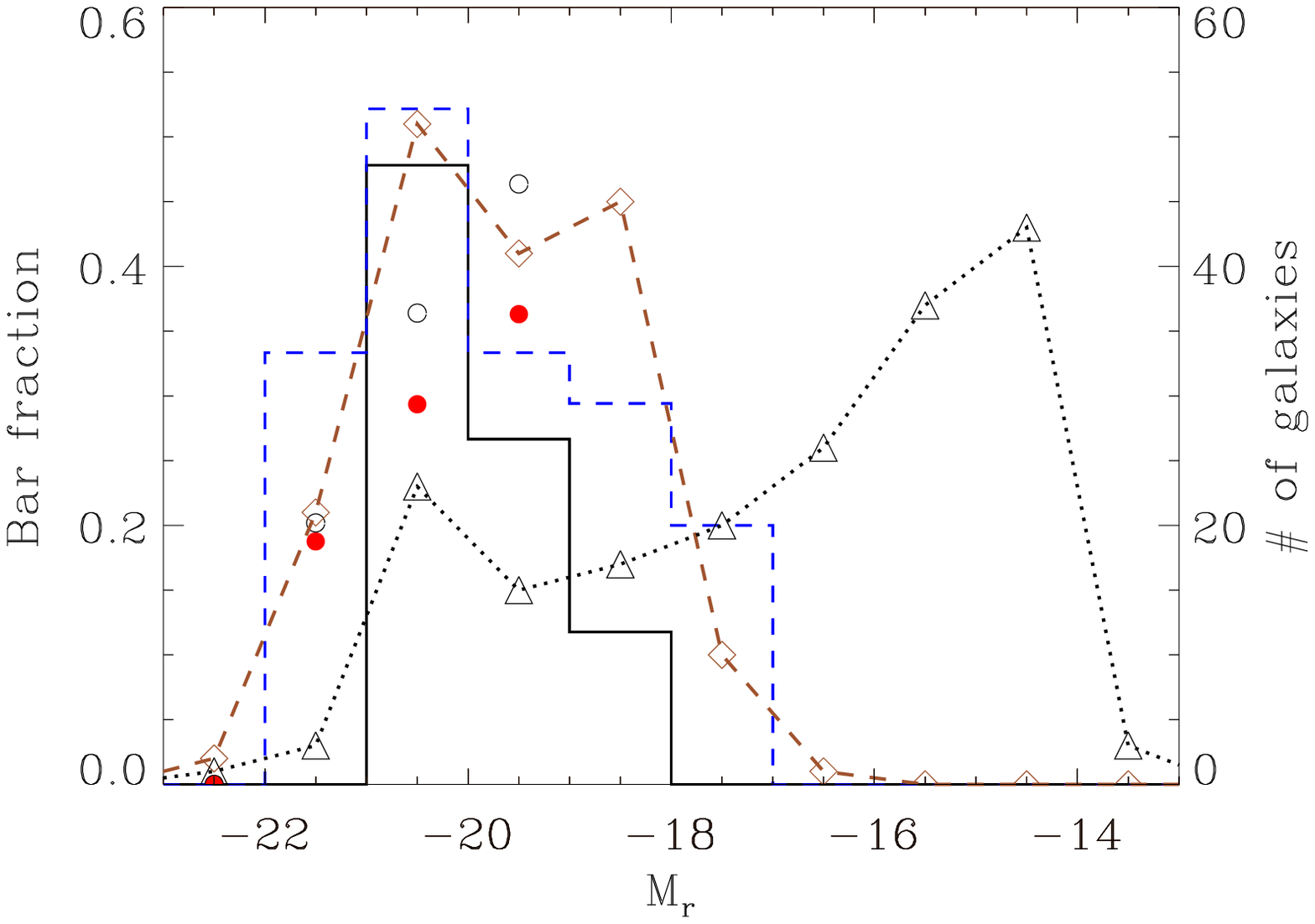}
  \includegraphics[width=0.49\textwidth]{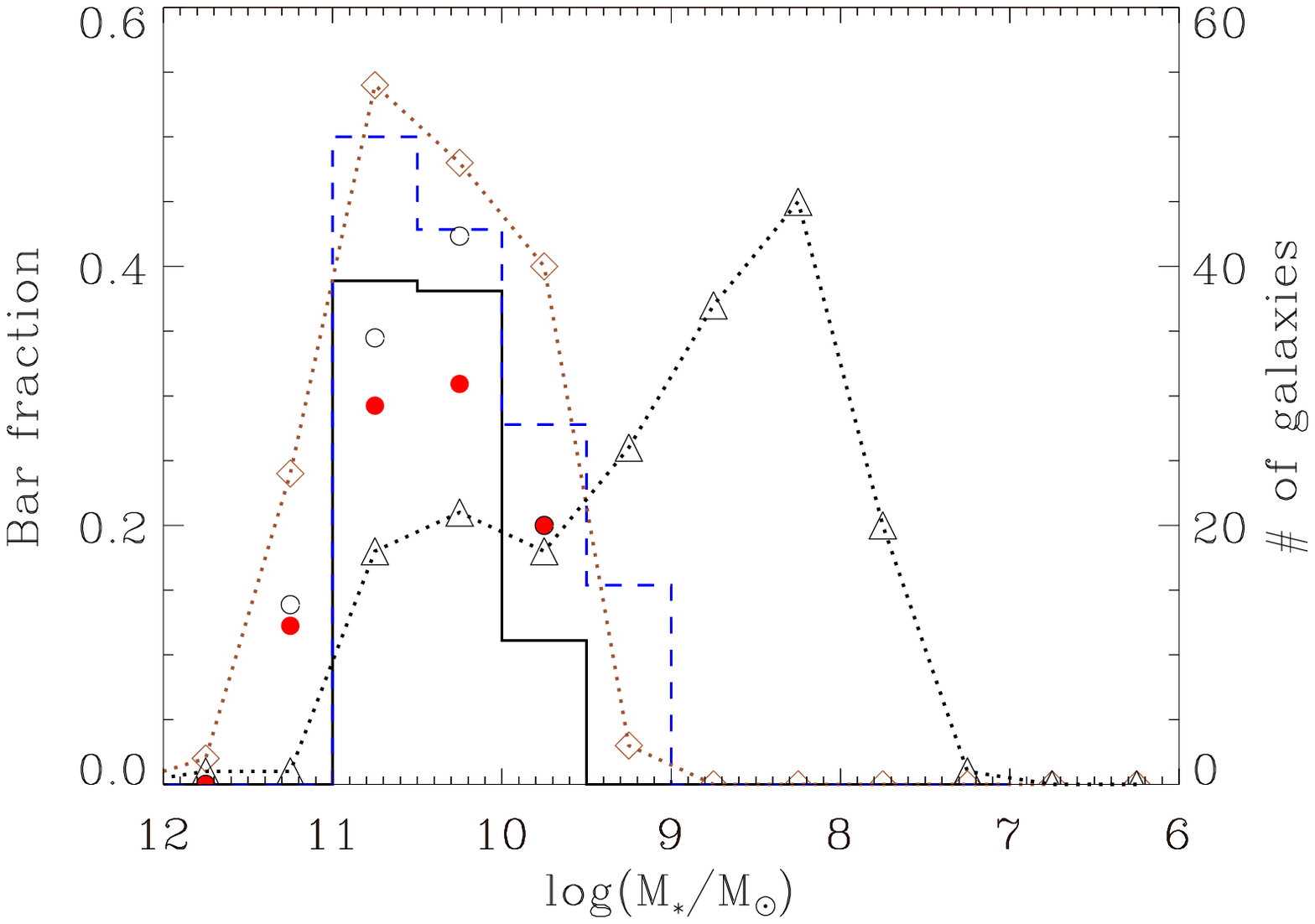}
  \caption{Optical  bar  fraction of  strong  (solid  black line)  and
    weak+strong  (dashed  blue  line)  as  a function  of  the  galaxy
    absolute  magnitude  in $r$-band  (left  panel),  and galaxy  mass
    (right panel).  Red points and black circles  represent the strong
    and   weak+strong  bar   fraction  using   the  field   sample  of
    \citet{aguerri09}, respectively. The number of galaxies per bin is
    represented   with   black   triangles.   The   disks   luminosity
    distribution  by  \citet{michardandreon08}  is  shown  with  brown
    diamonds,  the  $r$-band  magnitudes  were derived  from  the  $B$
    magnitudes by  using the  $B-R$ color for  every galaxy,  and then
    calculating  the $R$-$r$  differences by  taking into  account the
    equations  given by  Lupton  (2005). Stellar  masses were  derived
    following  the  prescriptions given  by  \citet{bell03} using  the
    $B-R$ color and the {\it diet} Salpeter IMF. }
              \label{fig:fractions}%
  \end{figure*}

In the present  work, we visually classified all  galaxies into strong
barred, weakly barred, and  unbarred. A caveat regarding this criteria
is that our  distinction between strong and weak  bars is not directly
related to the contribution of  the bar to the total galaxy potential,
but rather  they refer  to a  secure or possible  detection of  a bar,
respectively.   Therefore,   the  fraction  of  weak   bars  could  be
understood,  in some  way,  as a  measure  of our  uncertainty in  bar
detection. The visual classification was carried out by two of us (JMA
and  RSJ) using  the redder  available filter  (F814W) of  the HST-ACS
images.  Both  classifications were in close agreement  and only their
mean    is    reported   in    the    following    (see   also    Fig.
\ref{fig:fractions}).  Since  the goal of this paper  is to understand
where  do bars  form, we  have  not defined  the bar  fraction in  the
classic way, i.e., using only  disk galaxies, but we have instead used
all  galaxies   independently  of  their  Hubble   type.   Using  this
definition, our  bar fraction turns to  be $\sim 9\%$  and $\sim 14\%$
depending if only strong or also weak bars are included, respectively.

Fig. \ref{fig:fractions} shows  the bar fraction as a  function of the
luminosity and mass of the {\it secure} sample of galaxies in the Coma
cluster.  It is clear that independently of the bar strength, bars are
hosted by galaxies in a  tight range of luminosities or masses.  There
are no  strong bars in  the Coma cluster  out of the  luminosity range
between $-21  \lesssim $ M$_{r} \lesssim  -18$. These limits  become $-22
\lesssim $ M$_{r} \lesssim -17$ if we include also weak bars.

The  mass  of  the  galaxies  is one  of  the  fundamental  parameters
controlling  their  evolution.   We  use the  prescriptions  given  by
\citet{bell03} to  derive the stellar  mass of our galaxies  using the
$g-r$ color and the {\it diet} Salpeter initial mass function (IMF).
We find  again the same  behavior in the  bar fraction when  using the
galaxy  mass: bars  exist only  in a  range of  masses  between either
$10^{9.5} \lesssim {\cal M_{*}}/{\cal  M}_{\sun} \lesssim 10^{11} $ or
$10^{9}  \lesssim  {\cal   M_{*}}/{\cal  M}_{\sun}  \lesssim  10^{11}$
depending on  whether only strong or strong+weak  bars are considered,
respectively.

These findings rely on a relatively small number of galaxies and might
be affected by statistical  errors. Nevertheless, similar results were
found  when comparing  the  bright/massive side  of  our bar  fraction
distribution  with that  obtained, also  by visual  inspection  of the
galaxy  images, in  the  large  sample of  field  galaxies studied  by
\citet{aguerri09}, indicating that bars  are not hosted by very bright
galaxies (see Fig. \ref{fig:fractions}).

\section{Discussion and conclusions}
\label{sec:conclusions}

We have  used HST-ACS images taken  in the F814W filter  to search for
bars in  a sample of  188 galaxies members  of the Coma  cluster. Bars
were  identified  based  on  visual  inspection of  the  images.   The
unprecedented spatial  resolution provided  by HST-ACS in  this region
has allowed us to compute the bar fraction throughout a large range of
9 magnitudes,  permitting us  to explore the  presence of bars  in the
poorly known region of dwarf galaxies.

We find  that bars are  not hosted by  galaxies in the whole  range of
luminosities nor masses  covered in this study. On  the contrary, they
appear to be well constrained in a tight interval of both luminosities
$-22 \lesssim  $ M$_{r} \lesssim -17$  and masses $10^{9}  \lesssim {\cal
  M_{*}}/{\cal  M}_{\sun} \lesssim10^{11}$.   This result  has several
implications  on our  current  understanding of  bar formation  and/or
evolution.

If we assume that bars are tracers of cold stellar disks, the presence
of  bars is  particularly useful  to identify  galaxies with  disks in
clusters \citep{marinova09}.  Therefore, it  could be logical to think
that the distribution of bar fraction with the galaxy magnitude should
trace  the  shape  of   the  disk  galaxies  luminosity  distribution.
\citet{binggeli88}   showed  in  his   seminal  paper   the  magnitude
distribution of  morphological classes for  the Virgo cluster.  In the
bright  side of the  distribution ellipticals  are the  dominant type,
while both dwarf ellipticals and  dwarf irregulars are the most common
morphological  type  in  the  low  luminosity  region.  Instead,  disk
galaxies appear to  be constrained in the range  $-22 \lesssim $ M$_{B}
\lesssim -16$  in accordance with our  results. In order  to test this
hypothesis  in  the Coma  cluster,  we  have  computed the  luminosity
distribution of morphologically selected disk galaxies (from S0 to Sc)
using the classification carried out by \citet{michardandreon08}.  The
resulting   luminosity   distribution    (brown   diamonds   in   Fig.
\ref{fig:fractions})  matches  well  the  shape  of  the  bar  fraction
distribution  in both luminosity  and mass,  confirming that  bars are
good tracers of disks.

Since  simulations suggest that  bars form  spontaneously in  cold and
rotationally-supported disks,  the non  existence of bars  in galaxies
with M$_{r} \lesssim  -22$ can be interpreted either  as evidence that
cold disks do not form in such massive galaxies, or existing disks are
somewhat heated and they are not able to form/host a bar.

Disk heating should be predominant in high density environments due to
the higher frequency of  close encounters, major and minor accretions,
and the presence  of tidal forces, however, no  differences were found
between  field  and cluster  in  the  limiting  magnitude of  galaxies
hosting  bars.  This  constancy could  otherwise be  interpreted  as a
physical  limit   in  the  formation  of   disk  galaxies.   Numerical
simulations  carried  out  by \citet{dekel06}  and  \citet{cattaneo06}
suggest that  the physics  of the gas,  which will form  galaxy disks,
depends on the galaxy mass. They  claim that in halos below a critical
shock-heating mass  ${\cal M}_{\rm halo} \le  10^{12} {\cal M}_{\sun}$
disks are built  by cold gas streams while for  ${\cal M}_{\rm halo} >
10^{12} {\cal  M}_{\sun}$ the gas is  heated by a virial  shock and do
not  form  disks.  This   ${\cal  M}_{\rm  halo}  \sim  10^{12}  {\cal
  M}_{\sun}$  corresponds to  a  stellar mass  of  ${\cal M_{*}}  \sim
3\times10^{10}  {\cal  M}_{\sun}$  which  roughly coincides  with  our
limiting mass for galaxies hosting bars.

From  the environment  perspective  external triggers,  such as  tidal
interactions, can induce bars  \citep{noguchi87} but their effects can
be contradictory since  they may also heat the  disks and thereby make
them  less susceptible  to bar  formation.   It is  also important  to
consider  whether our  results  could  be explained  in  terms of  the
evolution  and   destruction  of   bars  rather  than   their  initial
formation. The scenario in which bars can be dissolved by the presence
of  a massive  central  concentration have  been  proposed in  several
simulations.   However,  most  of   them  indicate   that  present-day
supermassive black holes,  star clusters or inner parts  of bulges are
not     massive     enough     to    affect     bars     significantly
\citep[e.g.,][]{shensellwood04,  athanassoula05}.  On the  other hand,
if bars are hard to be dissolved once they are formed, the presence of
a  high  density environment  such  as  a  cluster should  not  change
dramatically the bar fraction, as we found in this work when comparing
with the field.

Since our sample galaxies cover  both the center and infall regions of
the Coma  cluster, we have further  tested this issue  by dividing our
sample  into internal and  external galaxies  and calculating  the bar
fraction for every  subsample. As we are introducing  large errors due
to small number statistics (especially  in the outer regions), we have
repeated this  procedure for three  values of the  separation distance
(0.5,  1, and  1.5 Mpc)  from the  cluster center  ($\alpha$: 12$^{h}$
59$^{m}$ 42$^{s}$,  $\delta$: 27$^{\circ}$ 58$^{'}$  15\farcs6, Godwin
et al.  1983).  For the smaller separation distance  we found 14\% and
15\% of  bars for the internal and  external subsamples, respectively.
The bar  fractions of both subsamples  are 14\% and 17\%  when using 1
Mpc, and 14\% and  17\% if we use 1.5 Mpc.  Therefore  we did not find
differences  in  the  bar  fraction  between the  subsamples  for  any
separation distance, implying again that the cluster environment plays
a second order role in bar formation/evolution.

In the  low luminosity/mass side  of the bar fraction  distribution we
found also a lack of  bars for galaxies with either M$_{r}\gtrsim -17$
or  ${\cal M_{*}}/{\cal  M}_{\sun} \lesssim  10^{9}$.  Few  works have
tried to investigate the dwarf realm in order to look for the presence
of  bars.   \citet{graham03}  found  spiral  structure  in  two  dwarf
galaxies belonging  to the Coma cluster. The  galaxies magnitudes were
found  to be  -18.8  and -17.4  in  the $R$-band,  therefore being  in
agreement with  our results.   \citet{lisker06}, studying a  sample of
dwarf  galaxies  in the  Virgo  cluster,  found  the presence  of  bar
structure in some dwarf galaxies as faint as M$_{B}\sim -16.10$, again
in full agreement with our  findings considering a typical $B-r$ color
$\sim1.8$.  They  also claim that dwarf ellipticals  with and without
disks  represent two  distinct types  of  galaxies, and  show how  the
fraction of dwarfs with  disk decrease dramatically for galaxies below
M$_{B}\sim -16$.  Therefore, even if the  non presence of  bars in our
low  luminosity galaxies  could  be due  to  the heating  of the  disk
component   or  to   its  absence,   our  results   support   that  of
\citet{lisker06}  and we  suggest that  no disks  are present  in Coma
galaxies below M$_{r}\sim -17$.

The physical mechanisms involved in  this case could be different with
respect  to  that  of  massive  galaxies.   The  role  played  by  the
environment  in the  evolution of  dwarfs is  crucial.   For instance,
repeated tidal  shocks suffered by  a dwarf satellite galaxy  can also
remove the kinematic  disk signature \citep{mayer01}. Cumulative tidal
fast encounters between galaxies  and with the gravitational potential
of   the  galaxy   cluster  can   produce  a   dramatic  morphological
transformation    from   spirals    to    roundish   dwarf    galaxies
\citep{moore96,mastropietro05,aguerrigonzalezgarcia09}.

We  are still  far from  understanding the  mechanisms that  drive one
galaxy to host a bar  while another apparently similar does not.  Even
if a great advance  has been done in the last years,  we still need to
explore the bunch of observational  data already available in order to
provide  more  inputs  to  numerical  simulations  to  understand  the
formation and evolutionary processes  of these important structures at
the center of galaxies.

\acknowledgments 

JMA   is  partially   funded   by  the   Spanish   MICINN  under   the
Consolider-Ingenio  2010 Program  grant  CSD2006-00070: First  Science
with the GTC  (http://www.iac.es/consolider-ingenio-gtc). JMA and JALA
are partially funded by the project AYA2007-67965-C03-01.  We thank V.
Debattista,  A.   de Lorenzo-Caceres,  and  I. Martinez-Valpuesta  for
useful  discussions  and  suggestions.   We also  thank  the  referee,
E. Laurikainen, for constructive  comments. Based on observations with
the NASA/ESA  Hubble Space Telescope  obtained at the STScI,  which is
operated by the association of Universities for Research in Astronomy,
Inc.,  under   NASA  contractNAS  5-26555.    These  observations  are
associated with program GO10861.


\begin{thebibliography}{}

\bibitem[Adelman-McCarthy et al.(2008)]{adelman08} 
Adelman-McCarthy, J.~K., et al.\ 2008, \apjs, 175, 297 
 
\bibitem[Aguerri et al.(1998)]{aguerri98} Aguerri, J.~A.~L., Beckman,
  J.~E., \& Prieto, M.\ 1998, \aj, 116, 2136

\bibitem[Aguerri et al.(2000)]{aguerri00} Aguerri, J.~A.~L.,
  Mu{\~n}oz-Tu{\~n}{\'o}n, C., Varela, A.~M., \& Prieto, M.\ 2000,
  \aap, 361, 841

\bibitem[Aguerri    et   al.(2009)]{aguerri09}    Aguerri,   J.~A.~L.,
  M{\'e}ndez-Abreu, J., \& Corsini, E.~M.\ 2009, \aap, 495, 491

\bibitem[Aguerri                                                     \&
  Gonz{\'a}lez-Garc{\'{\i}}a(2009)]{aguerrigonzalezgarcia09}
  Aguerri, J.~A.~L., \& Gonz{\'a}lez-Garc{\'{\i}}a, A.~C.\ 2009, \aap,
  494, 891

\bibitem[Athanassoula(2003)]{athanassoula03} Athanassoula, E.\ 2003,
  \mnras, 341, 1179

 \bibitem[Athanassoula et al.(2005)]{athanassoula05} Athanassoula, E., 
Lambert, J.~C., \& Dehnen, W.\ 2005, \mnras, 363, 496 

\bibitem[Barazza et al.(2008)]{barazza08} Barazza, F.~D., Jogee, S.,
  \& Marinova, I.\ 2008, \apj, 675, 1194

\bibitem[Bell et al.(2003)]{bell03} Bell, E.~F., McIntosh, 
D.~H., Katz, N., \& Weinberg, M.~D.\ 2003, \apjs, 149, 289 

\bibitem[Binggeli  et  al.(1988)]{binggeli88}  Binggeli,  B.,
  Sandage, A., \& Tammann, G.~A.\ 1988, \araa, 26, 509

\bibitem[Bureau \&  Freeman(1999)]{bureaufreeman99} Bureau, M.,  \& Freeman,
  K.~C.\ 1999, \aj, 118, 126
 
\bibitem[Buta et al.(2003)]{buta03} Buta, R., Block, D.~L., \& Knapen,
  J.~H.\ 2003, \aj, 126, 1148
 
\bibitem[Carter et al.(2008)]{carter08} Carter, D., et al.\ 
2008, \apjs, 176, 424

\bibitem[Cattaneo et al.(2006)]{cattaneo06} Cattaneo, A., Dekel, 
A., Devriendt, J., Guiderdoni, B., \& Blaizot, J.\ 2006, \mnras, 370, 1651 

\bibitem[Colless \&  Dunn(1996)]{collessdunn96} Colless, M.,  \& Dunn,
  A.~M.\ 1996, \apj, 458, 435

\bibitem[Corsini et al.(2003)]{corsini03} Corsini, E.~M., Debattista,
  V.~P., \& Aguerri, J.~A.~L.\ 2003, \apjl, 599, L29

\bibitem[Debattista \& Sellwood(1998)]{debattistasellwood98} Debattista,
  V.~P., \& Sellwood, J.~A.\ 1998, \apjl, 493, L5
  
\bibitem[Debattista \& Sellwood(2000)]{debattistasellwood00} Debattista,
  V.~P., \& Sellwood, J.~A.\ 2000, \apj, 543, 704

\bibitem[Debattista  \&  Shen(2007)]{debattistashen07}  Debattista,
  V.~P., \& Shen, J.\ 2007, \apjl, 654, L127

\bibitem[Dekel  \& Birnboim(2006)]{dekel06}  Dekel,  A., \&  Birnboim,
  Y.\ 2006, \mnras, 368, 2

\bibitem[de Vaucouleurs et al.(1991)]{devaucouleurs91} de Vaucouleurs,
  G., de Vaucouleurs, A., Corwin, H.\ G.\ Jr., Buta, R.\ J., Paturel,
  G., \& Fouqu\`e, P.  1991, Third Reference Catalogue of Bright
  Galaxies (Berlin: Springer-Verlag)

\bibitem[Elmegreen \& Elmegreen(1985)]{elmegreenelmegreen85} Elmegreen, B.~G.,
  \& Elmegreen, D.~M.\ 1985, \apj, 288, 438

\bibitem[Erwin(2004)]{erwin04} Erwin, P.\ 2004, \aap, 415, 941

\bibitem[Godwin et al.(1983)]{godwin83} Godwin, J.~G., Metcalfe, 
N., \& Peach, J.~V.\ 1983, \mnras, 202, 113 

\bibitem[Graham et al.(2003)]{graham03} Graham, A.~W., Jerjen, 
H., \& Guzm{\'a}n, R.\ 2003, \aj, 126, 1787 

\bibitem[Hohl(1971)]{hohl71} Hohl, F.\ 1971, \apj, 168, 343 
 
\bibitem[Kormendy \& Kennicutt(2004)]{kormendykennicutt04} Kormendy,
  J., \& Kennicutt, R.~C., Jr.\ 2004, \araa, 42, 603
 
\bibitem[Laurikainen et al.(2004)]{laurikainen04} Laurikainen, E., 
Salo, H., \& Buta, R.\ 2004, \apj, 607, 103 

\bibitem[Laurikainen et al.(2005)]{laurikainen05} Laurikainen, E.,
  Salo, H., \& Buta, R.\ 2005, \mnras, 362, 1319
  
\bibitem[Lisker et al.(2006)]{lisker06} Lisker, T., Grebel, 
E.~K., \& Binggeli, B.\ 2006, \aj, 132, 497 

\bibitem[Marinova \& Jogee(2007)]{marinovajogee07} Marinova, I., \& Jogee,
  S.\ 2007, \apj, 659, 1176
  
\bibitem[Marinova   et   al.(2009)]{marinova09}   Marinova,   I.,   et
  al.\ 2009, \apj, 698, 1639

\bibitem[Martinez-Valpuesta et al.(2006)]{martinezvalpuesta06}
  Martinez-Valpuesta, I., Shlosman, I., \& Heller, C.\ 2006, \apj,
  637, 214

\bibitem[Mastropietro et al.(2005)]{mastropietro05} Mastropietro, C.,
  Moore, B., Mayer, L., Debattista, V.~P., Piffaretti, R., \& Stadel,
  J.\ 2005, \mnras, 364, 607

\bibitem[Mayer et al.(2001)]{mayer01} Mayer, L., Governato, F., 
Colpi, M., Moore, B., Quinn, T., Wadsley, J., Stadel, J., 
\& Lake, G.\ 2001, \apj, 559, 754 

\bibitem[M{\'e}ndez-Abreu et al.(2008)]{mendezabreu08} 
M{\'e}ndez-Abreu, J., Corsini, E.~M., Debattista, V.~P., De Rijcke, S., 
Aguerri, J.~A.~L., \& Pizzella, A.\ 2008, \apjl, 679, L73 

\bibitem[Men{\'e}ndez-Delmestre et al.(2007)]{menendezdelmestre07}
  Men{\'e}ndez-Delmestre, K., Sheth, K., Schinnerer, E., Jarrett,
  T.~H., \& Scoville, N.~Z.\ 2007, \apj, 657, 790

\bibitem[Michard  \& Andreon(2008)]{michardandreon08} Michard,  R., \&
  Andreon, S.\ 2008, \aap, 490, 923

\bibitem[Moore et al.(1996)]{moore96} Moore, B., Katz, N.,
  Lake, G., Dressler, A., \& Oemler, A.\ 1996, \nat, 379, 613

\bibitem[Mu{\~n}oz-Tu{\~n}{\'o}n   et  al.(2004)]{munoztunon04}
  Mu{\~n}oz-Tu{\~n}{\'o}n, C.,  Caon, N., \&  Aguerri, J.~A.~L.\ 2004,
  \aj, 127, 58

\bibitem[Noguchi(1987)]{noguchi87} Noguchi, M.\ 1987, \mnras, 228, 635

\bibitem[Ohta et al.(1990)]{ohta90} Ohta, K., Hamabe, M., \&
  Wakamatsu, K.-I.\ 1990, \apj, 357, 71

\bibitem[Ostriker   \&   Peebles(1973)]{ostrikerpeebles73}   Ostriker,
  J.~P., \& Peebles, P.~J.~E.\ 1973, \apj, 186, 467

\bibitem[Sellwood \& Debattista(2006)]{sellwooddebattista06} Sellwood,
  J.~A., \& Debattista, V.~P.\ 2006, \apj, 639, 868

\bibitem[Shen  \&  Sellwood(2004)]{shensellwood04}  Shen, J.,  \&
  Sellwood, J.~A.\ 2004, \apj, 604, 614

\bibitem[Sheth et al.(2005)]{sheth05} Sheth, K., Vogel, S.~N., 
Regan, M.~W., Thornley, M.~D., \& Teuben, P.~J.\ 2005, \apj, 632, 217 

\bibitem[Sheth et al.(2008)]{sheth08} Sheth, K., et al.\ 2008, 
\apj, 675, 1141

\bibitem[Shlosman et al.(2000)]{shlosman00} Shlosman, I., Peletier,
  R.~F., \& Knapen, J.~H.\ 2000, \apjl, 535, L83

\bibitem[Weinzirl et al.(2009)]{weinzirl09} Weinzirl, T., Jogee, 
S., Khochfar, S., Burkert, A., \& Kormendy, J.\ 2009, \apj, 696, 411 
 
\end{thebibliography}
\end{document}